# A Discrepancy in Thermal Conductivity Measurement Data of Quantum Spin Liquid β′-EtMe$_3$Sb[Pd(dmit)$_2$]$_2$ (dmit = 1,3-Dithiol-2-thione-4,5-dithiolate)

Reizo Kato *, Masashi Uebe, Shigeki Fujiyama and Hengbo Cui

Condensed Molecular Materials Laboratory RIKEN, Wako, Saitama 351-0198, Japan;
masaliam020902@gmail.com (M.U.); fujiyama@riken.jp (S.F.); hengbocui@gmail.com (H.C.)
* Correspondence: reizo@riken.jp

**Abstract:** A molecular Mott insulator β′-EtMe$_3$Sb[Pd(dmit)$_2$]$_2$ is a quantum spin liquid candidate. In 2010, it was reported that thermal conductivity of β′-EtMe$_3$Sb[Pd(dmit)$_2$]$_2$ is characterized by its large value and gapless behavior (a finite temperature-linear term). In 2019, however, two other research groups reported opposite data (much smaller value and a vanishingly small temperature-linear term) and the discrepancy in the thermal conductivity measurement data emerges as a serious problem concerning the ground state of the quantum spin liquid. Recently, the cooling rate was proposed to be an origin of the discrepancy. We examined effects of the cooling rate on electrical resistivity, low-temperature crystal structure, and $^{13}$C-NMR measurements and could not find any significant cooling rate dependence.

## 1. Introduction

Quantum spin liquid (QSL) in a strongly frustrated spin system on a triangular lattice is characterized by the absence of long-range magnetic order or valence bond solid order among entangled quantum spins even at zero temperature [1,2]. Although theoretical works indicated that the ideal nearest-neighbor Heisenberg antiferromagnet on the triangular lattice has the long-range Néel ordered ground state (120-degree-structured state), a possibility of this third fundamental state for magnetism in the general $S = 1/2$ antiferromagnetic triangular lattice systems has attracted much attention. Indeed, the number of QSL candidates of real materials is increasing since the beginning of the 2000 [2].

An isostructural series of anion radical salts of a metal complex Pd(dmit)$_2$ (dmit = 1,3-dithiol-2-thione-4,5-dithiolate), β′-Et$_x$Me$_{4-x}$Z[Pd(dmit)$_2$]$_2$ (Et: C$_2$H$_5$-, Me: CH$_3$-, Z = P, As, and Sb; $x$ = 0–2), are Mott insulators at ambient pressure [3]. In crystals of β′-Pd(dmit)$_2$ salts with the space group $C2/c$, Pd(dmit)$_2$ anion radicals are strongly dimerized to form a dimer with spin 1/2, [Pd(dmit)$_2$]$_2^-$ (Figure 1). The dimers are arranged in an approximately isosceles-triangular lattice parallel to the *ab* plane, which leads to a frustrated $S = 1/2$ Heisenberg spin system. The anion radical layers and the non-magnetic cation layers are arranged alternately along the *c* axis. The ground state of the Pd(dmit)$_2$ salts is found to change among antiferromagnetic long-range order (AFLO), QSL, and charge order (CO), depending on the anisotropy of the triangular lattice that can be tuned by the choice of the counter cation [3]. The cation effect on the degree of frustration is associated with the arch-shaped distortion of the Pd(dmit)$_2$ molecule [4]. The QSL phase found in β′-EtMe$_3$Sb[Pd(dmit)$_2$]$_2$ is situated between AFLO and CO phases [5,6].

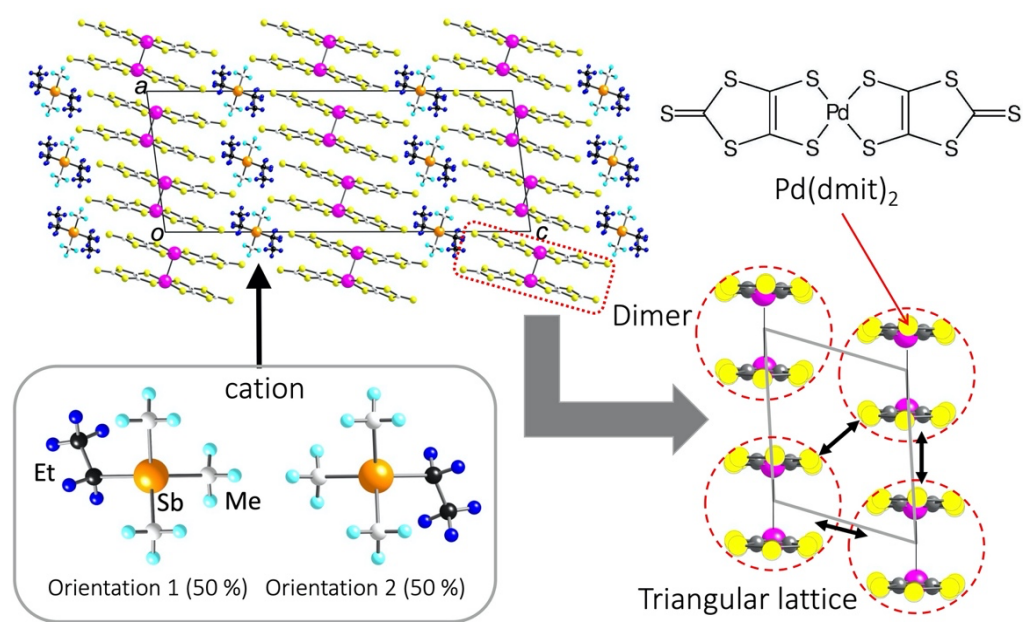

**Figure 1.** Crystal structure of β′-EtMe$_3$Sb[Pd(dmit)$_2$]$_2$. The cation on the two-fold axis shows an orientational disorder.

In β′-EtMe$_3$Sb[Pd(dmit)$_2$]$_2$, no magnetic order is detected down to a very low temperature (~19 mK) that corresponds to $J/12,000$, where $J$ (~250 K) is the nearest-neighbor spin interaction energy [7]. Although $^{13}$C-NMR spectra show an inhomogeneous broadening at low temperatures, the observed local static fields are too small to be explained by AFLO or spin glass state. Low-energy excitations in the QSL state of the β′-EtMe$_3$Sb[Pd(dmit)$_2$]$_2$ are open to debate even now. Heat capacity and magnetization indicate gapless fermion-like excitations, while $^{13}$C-NMR indicates an existence of a nodal gap [7–9]. The thermal conductivity $\kappa$ is free from the contribution from the rotation of the methyl (Me-) group that disturbs the heat capacity analysis below 1 K [8]. In addition, thermal conductivity measurements can detect the spin-mediated heat transport. In 2010, Yamashita and Matsuda reported that $\kappa/T$ is finite as temperature $T$ goes to zero, which indicates the presence of gapless excitations [10]. The finite $T$-linear term as well as largely enhanced $\kappa$ values led to a proposal of contributions across the spinon Fermi surface. In 2019, however, two research groups reported that $\kappa$ values are much smaller and $\kappa/T$ is vanishingly small at 0 K, which caused a serious problem concerning the ground state of QSL [11,12].

In order to explain this sharp discrepancy in the thermal conductivity measurement data, Yamashita claimed that there were two kinds of crystals (large-$\kappa$ and small-$\kappa$ groups) in [13] published earlier than [11,12]. Yamashita pointed out the domain formation associated with the cation disorder or the micro cracks as an origin. It should be noted that in the context of "two kinds of crystals", the words "domain" and "micro cracks" are read as intrinsic properties that emerge in a crystal growth process or in a low-temperature phase, that is, they should be distinguished from extrinsic ones induced by improper sample handling. Although Yamashita did not disclose experimental evidence to justify the claim in [13], the claim had enough impact [14,15]. In response to the Yamashita's claim, the existence of two kinds of crystals was verified using X-ray diffraction (XRD), scanning electron microscope, and electrical resistivity measurements. The conclusion is that there is only one kind of crystal [11,12]. For example, no difference was found between the small-$\kappa$ sample (sample G1) in [11] and the large-$\kappa$ sample (sample C) in [13], both of which come from the very same growth batch (No. 752).

Meanwhile, in 2020, Yamashita et al. reported that one kind of crystal gives different results in the $\kappa$ measurements depending on an experimental condition, "cooling rate" [16,17]. In their measurements, very slow cooling (−0.4 K/h) led to a finite linear residual thermal conductivity. In contrast, when the sample was cooled down rapidly (−13 K/h), $\kappa/T$ vanished at the zero-temperature limit, and the phonon thermal conductivity was strongly suppressed. These results suggest the existence of random scatterers introduced during the cooling process as another origin of the discrepancy. This proposal has raised a problem about effects of the newly proposed experimental parameter on other kinds of measurements. Herein, we investigated effects of the cooling rate on electrical resistivity, low-temperature crystal structure, and $^{13}$C-NMR measurements with relevance to the discrepancy in the thermal conductivity data.

## 2. Materials and Methods

For electrical resistivity and XRD measurements, we used single crystals from the same growth batch (No. 898) as that used in [13] (small-$\kappa$ samples E and F) and [16] (crystals 1–3). The procedure of the crystal growth was as follows: (EtMe$_3$Sb)$_2$[Pd(dmit)$_2$] (60 mg) was dissolved in acetone (100 mL). After addition of acetic acid (9.5 mL), the resultant solution was allowed to stand at −11 °C for 3 months. The β′-type crystals (black hexagonal plates) were obtained as a single phase. The $^{13}$C-enriched dmit ligand for $^{13}$C-NMR measurements was synthesized from tetrachloroethylene-$^{13}$C$_1$ (99 atom %, Sigma-Aldrich) that was converted to tetrathiooxalate for the reaction with CS$_2$ [18]. The $^{13}$C-enriched single crystals of β′-EtMe$_3$Sb[Pd(dmit)$_2$]$_2$ (No. 899 for slow cooling and No. 923 for rapid cooling) were also obtained by the above-mentioned procedure.

The temperature-dependent electrical resistivities along the $a$ and $b$ axes ($\rho_{//a}$ and $\rho_{//b}$) were measured by the standard four-probe method from room temperature to 1.8 K using a physical property measurement system (Quantum Design Inc., San Diego, CA, USA). The electric leads were $\phi$10 μm gold wires connected by carbon paste. Probe sizes are 325 × (1100 × 100) μm$^3$ for $\rho_{//a}$ and 150 × (1350 × 80) μm$^3$ for $\rho_{//b}$, respectively. For each current direction, the pristine crystal was cooled down to 1.8 K with different cooling rates of −0.6 K/h, −1.2 K/h, −6 K/h, and −150 K/h, in this order. In each thermal cycle, the warming rate was +6 K/h, except for the final cycle (+150 K/h).

Single crystal X-ray diffraction data were collected by a Weisenberg-type imaging plate system (R-AXIS RAPID/CS, Rigaku Corp., Tokyo, Japan) with monochromated Mo K$\alpha$ radiation (UltraX6-E, Rigaku Corp., Tokyo, Japan). Low-temperature experiments were carried out in the cryostat cooled by a closed-cycle helium refrigerator (HE05/UV404, ULVAC CRYOGENICS Inc., Chigasaki, Japan). The temperature was controlled by Model 22C Cryogenic Temperature Controller (Cryogenic Control Systems, Inc., Rancho Santa Fe, CA, USA). The pristine crystal was cooled down with a rate of −0.6 K/h. After the data collection at 5 K, the crystal was warmed up to room temperature with a rate of +60 K/h. The next cooling process was performed with a cooling rate of −120 K/h. All the diffraction data were processed using the CrystalStructure 3.8 crystallographic software package [19]. The structures were solved by the direct method (SIR92) [20] and refined by the full-matrix least-squares method (SHELXL-2018/3) [21]. The H atom coordinates were placed on calculated positions and refined with the riding model. Due to the orientational disorder of the EtMe$_3$Sb cation on the two-fold axis (Figure 1), we assumed that the ethyl group and the corresponding methyl group share two equivalent positions with 50% occupation factor for the refinements.

$^{13}$C-NMR spectra and nuclear relaxation rates were obtained by standard pulse Fourier transform technique using a single crystal. The magnetic field was applied along the direction 11 degrees tilted from the $c^*$ axis to avoid an accidental cancellation of hyperfine fields originating from $2p$ and $2s$ electrons of $^{13}$C. The temperature was controlled by Model 32 Cryogenic Temperature Controller. The crystals were cooled at −40 K/h (rapid cooling) and −0.6 K/h (slow cooling), respectively.

## 3. Results

*3.1. Electrical Resistivity*

The electrical resistivity is sensitive to the crack formation. In addition, when the emergence of an electronic phase depends on the cooling rate as is the case of θ-(BEDT-TTF)$_2$RbZn(SCN)$_4$ (BEDT-TTF = Bis(ethylenedithio)tetrathiafulvalene) that possesses the charge-glass-forming ability, the electrical resistivity can detect a change of the electronic state [22]. Figure 2 shows temperature-dependent resistivities along the *a* and *b* axes measured with four different cooling rates, −0.6, −1.2, −6, and −150 K/h in this order. β'-EtMe$_3$Sb[Pd(dmit)$_2$]$_2$ was a semiconductor and the resistivity became too high to measure below 28 K. Anisotropy within the *ab* plane was small, including the activation energy (~41 meV). As shown in Figure 2, temperature-resistivity curves for each cooling rate overlap almost completely, which means that there was no cooling rate dependence. In addition, no crack formation (indicated by an abrupt jump of the resistivity) and no thermal cycle dependence was observed.

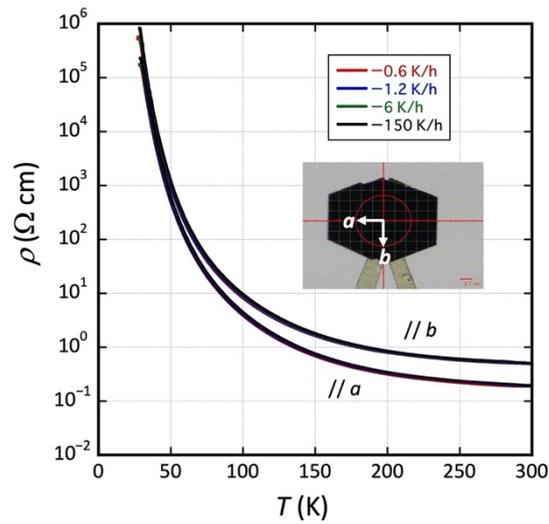

**Figure 2.** Temperature-dependent resistivities measured along the *a* axis ($\rho_{//a}$) and the *b* axis ($\rho_{//b}$) with four different cooling rates for β'-EtMe$_3$Sb[Pd(dmit)$_2$]$_2$. The inset is a photo of a single crystal. For both $\rho_{//a}$ and $\rho_{//b}$, four $\rho$–$T$ curves (cooling process) overlap almost completely.

*3.2. Low-Temperature Crystal Structure*

Using the same single crystal, the crystal structure of β'-EtMe$_3$Sb[Pd(dmit)$_2$]$_2$ at 5 K was determined with two different cooling rates, −0.6 (slow) and −120 (rapid) K/h in this order. In both cases, the space group remained *C*2/*c* and no additional diffraction peak was observed. Determined crystal structures were identical with the previous result [23], and did not show any significant effects of the cooling rate on temperature factors and differential Fourier synthesis (Table 1).

**Table 1.** Crystal data for β'-EtMe$_3$Sb[Pd(dmit)$_2$]$_2$ at 5 K obtained using two different cooling rates.

| Cooling Rate | −120 K/h (Rapid) | −0.6 K/h (Slow) |
|---|---|---|
| *a* (Å) | 14.346 (8) | 14.342 (8) |
| *b* (Å) | 6.317 (3) | 6.315 (3) |
| *c* (Å) | 37.07 (2) | 37.07 (2) |
| β (Å) | 97.624 (14) | 97.640 (13) |
| *V* (Å$^3$) | 3330 (3) | 3328 (3) |
| R factor | 0.0370 | 0.0357 |
| G.O.F. | 1.079 | 1.049 |
| Δ$\rho_{max}$ (e Å$^{-3}$) | 1.363 | 1.314 |
| Δ$\rho_{min}$ (e Å$^{-3}$) | −1.089 | −0.960 |

In the crystal, the EtMe₃Sb⁺ cation is located on the 2-fold axis (//b). Since the cation does not have the 2-fold symmetry, the cation has two possible orientations (1 and 2 in Figure 1), which could work as an origin of the domain formation. However, our analysis, where the ethyl group is assumed to be overlaid with the methyl group with 50% occupation factor, did not find significant difference in the average cation structure for both cooling rates (Figure 3).

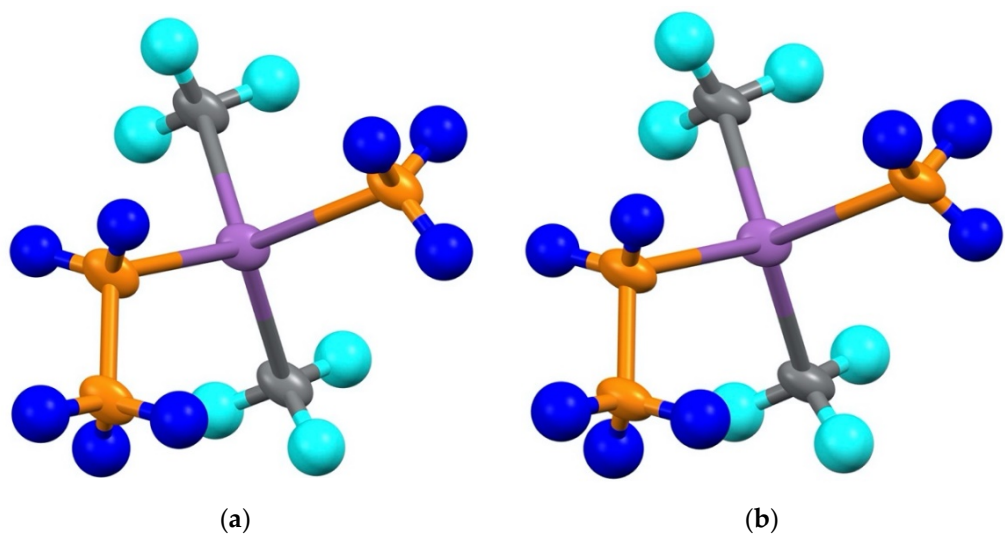

(**a**) (**b**)

**Figure 3.** Average structure of the cation at 5 K viewed from the $b$ axis (the two-fold axis) for two different cooling rates: (**a**) Rapid cooling (−120 K/h); and (**b**) slow cooling (−0.6 K/h).

*3.3. ¹³C-NMR*

¹³C-NMR enables us to investigate the microscopic electronic states of a crystal. Compared with the electrical resistivity, both NMR spectra and nuclear relaxations are insensitive to microcracks in a crystal. However, they would be able to detect possible domain formations in a cooling procedure.

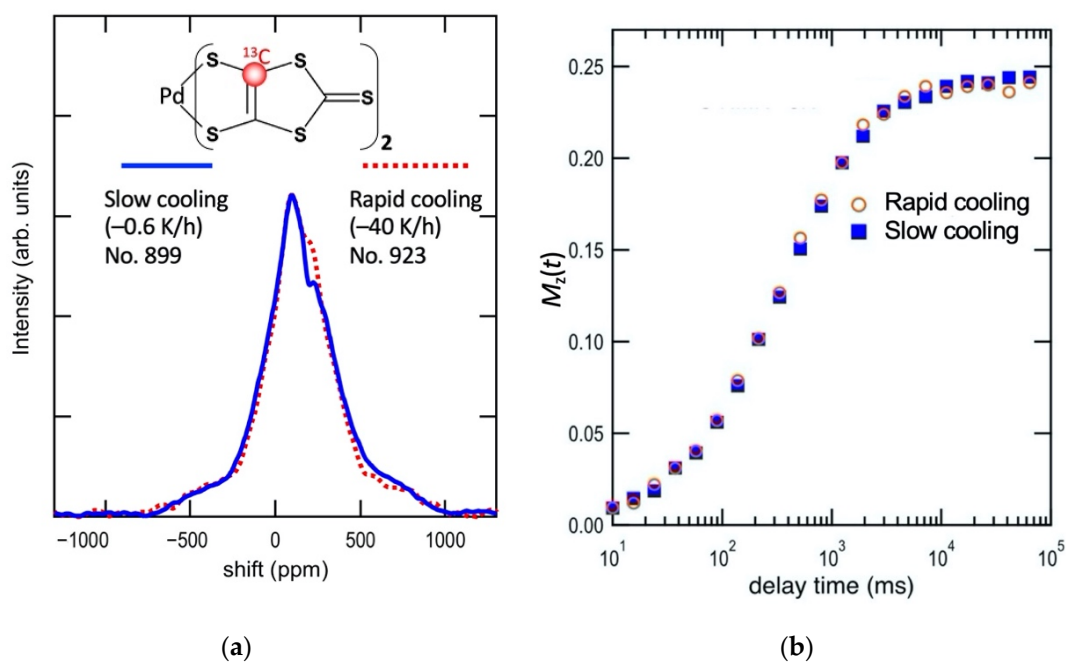

(**a**) (**b**)

**Figure 4.** ¹³C-NMR of β′-EtMe₃Sb[Pd(dmit)₂]₂ at 5 K with rapid and slow cooling rates. (**a**) Spectra; and (**b**) nuclear magnetization curves.

Figure 4a shows the spectra at 5K. The $^{13}$C atom is introduced into one of the carbon sites in the C=C bond in the dmit ligand. The four independent $^{13}$C sites at which hyperfine coupling constants distribute up to 10% cause asymmetric spectra [24]. Two spectra with contrasting cooling rates nearly overlap with each other, by which we conclude negligible cooling rate variation in the static electronic states.

Nuclear magnetization ($M_z$) curves of the $1/T_1$ (nuclear spin-lattice relaxation rate) measurements at 5 K by the rapid and slow cooling procedures are shown in Figure 4b. Two curves agree with each other, and we conclude that the spin dynamics of the quantum spin liquid state is insensitive to the cooling rate.

It should be noted that the previous $^{13}$C-NMR data obtained using randomly orientated crystals suggest no cooling rate dependence either [5,7]. In [5], the cooling process at a rate of ca. −10 K/h and measurements at a constant temperature (during 1~2 days) were repeated alternately, and the sample was cooled from room temperature down to 1.4 K spending about one month. In [7], on the other hand, the sample was cooled from room temperature down to 1.8 K within 10 h before the measurements in very low-temperature region (1.8 K–20 mK). The $^{13}$C-NMR data with these two different experimental conditions coincide with each other in the same temperature region [25].

## 4. Discussion

In this work, we could not observe the effect of the cooling rate on resistivity, low-temperature crystal structure, and $^{13}$C-NMR. Of course, we must be careful in discussing the relation between these physical/structural properties and the thermal conductivity. The problem we are facing is the thermal conductivity below 1 K. In the low temperature region, the electrical resistivity of the present material is very high, and the mean free path of a charge carrier becomes shorter than lattice lengths. In such a case, conduction electrons would be unaffected by an event in the whole crystal. On the other hand, the low-temperature crystal structures we determined are average ones and do not provide direct information about the possible domain or defect formation.

Nevertheless, the cooling rate engages in a process in the whole temperature region. If random scatterers are generated during the cooling process, they would be detected by physical/structural properties other than the thermal conductivity even in the high temperature region. In addition, local changes in a crystal could affect an average crystal structure and $^{13}$C-NMR. As we mentioned before, the cooling rate dependence was not detected in the high temperature region (> ~5 K) in this work. In the lower temperature region, the smaller heat capacity provides more homogeneous temperature distribution, and thus it is less plausible that the cooling rate plays an important role.

In this sense, the results of this work suggest that further analysis is necessary before concluding that the cooling rate is an essential experimental condition. Let us reconsider two different sets of thermal conductivity data from [10,16] (Figure 5). Measurements with different cooling rates indicated that the slower cooling rate gave the larger $\kappa$ and finite $\kappa/T$ [16]. However, even with the slowest cooling rate (−0.4 K/h), the $\kappa$ values are much smaller than the first reported ones [10]. That is, the large-$\kappa$ data have never been reproduced. In addition, the measurements in [10] performed with the rapid cooling rate of −10 K/h show larger $\kappa$ values than those for Crystal 1 measured with the slowest cooling rate of −0.4 K/h in [16]. This is quite puzzling and suggests that the cooling rate is not essential.

In conclusion, the present situation is that one kind of crystal provides two different thermal conductivity data and a role of the newly proposed experimental parameter, cooling rate, remains to be seen. In addition, the large-$\kappa$ data have never been reproduced. It is an urgent matter to clarify the intrinsic thermal conductivity. From this point of view, the description "*the crystals often do not recover to the initial state after a thermal cycle*" in [16] suggests that the stress from experimental environments including a setup may enhance or suppress the thermal conductivity. Indeed, crystals used in [10,16] frequently fell apart when the leads were removed by rinsing out the paste with diethyl succinate. This suggests the existence of the stress from leads on a crystal. In contrast, we did not observe any thermal cycle dependence in this work. In order to clarify this point, monitoring of electrical resistivity during thermal conductivity measurements will be valuable, because the resistivity is sensitive to the crack formation and pressure in the wide range of temperature (> ~28K).

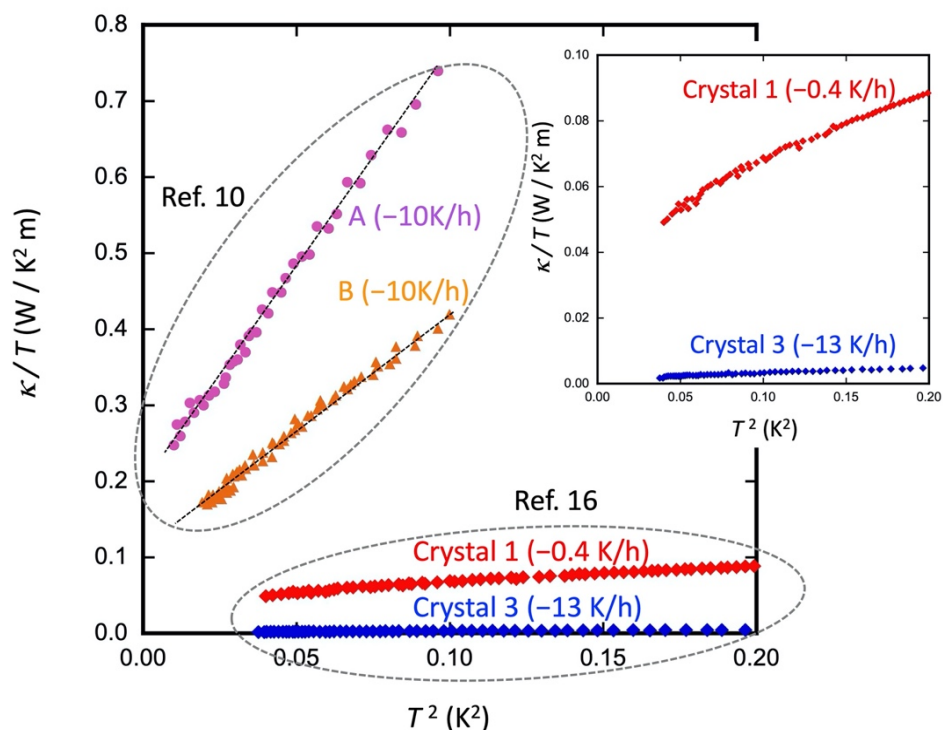

**Figure 5.** Two different sets of the thermal conductivity ($\kappa$) data for β'-EtMe$_3$Sb[Pd(dmit)$_2$]$_2$ from [10,16]. The cooling rate in each measurement is indicated in a parenthesis. The inset is an enlarged view of the data from [16]. The behavior of $\kappa$ reported in [11,12] is similar to that of crystal 3 (−13 K/h) in [16].

**Author Contributions:** Conceptualization, funding acquisition, project administration, sample preparation, and writing—original draft preparation, R.K.; investigation and data curation, M.U., S.F. and H.C. All authors have read and agreed to the published version of the manuscript.

**Funding:** This research was partially supported by JSPS KAKENHI [grant number JP16H06346].

**Institutional Review Board Statement:** Not applicable.

**Informed Consent Statement:** Not applicable.

**Data Availability Statement:** Crystallographic information files are available from the CCDC, reference numbers 2125994-2125995. These data can be obtained free of charge via https://www.ccdc.cam.ac.uk/structures/.

**Acknowledgments:** We deeply thank Dr. Daisuke Hashizume (RIKEN Center for Emergent Matter Science) for technical help with the crystal structure analysis of β'-EtMe$_3$Sb[Pd(dmit)$_2$]$_2$ at 5 K.

**Conflicts of Interest:** The authors declare no conflict of interest.

**References**

1. Anderson, P.W. Resonating valence bonds: A new kind of insulator? *Mater. Res. Bull.* **1973**, *8*, 153–160.
2. Balents, L. Spin liquids in frustrated magnets. *Nature* **2010**, *464*, 199–208.
3. Kato, R. Development of π-electron systems based on [M(dmit)$_2$] (M= Ni and Pd; dmit: 1,3-dithiole-2-thione-4,5-dithiolate) Anion Radicals. *Bull. Chem. Soc. Jpn.* **2014**, *87*, 355–374.

4. Kato, R.; Hengbo, C. Cation dependence of crystal structure and band parameters in a series of molecular conductors, β'-(Cation)[Pd(dmit)$_2$]$_2$ (dmit = 1,3-dithiol-2-thione-4,5-dithiolate). *Crystals* **2012**, *2*, 861–874.
5. Itou, T; Oyamada, A; Maegawa, S.; Tamura, M.; Kato, R. Quantum spin liquid in the spin-1/2 triangular antiferromagnet EtMe$_3$Sb[Pd(dmit)$_2$]$_2$. *Phys. Rev. B*. **2008**, *77*, 104413.
6. Kanoda, K.; Kato, R. Mott physics in organic conductors with triangular lattices. *Annu. Rev. Condens. Matter Phys*. **2011**, *2*, 167–188.
7. Itou, T; Oyamada, A; Maegawa, S.; Kato, R. Instability of a quantum spin liquid in an organic triangular-lattice antiferromagnet. *Nat. Phys*. **2010**, *6*, 673–676.
8. Yamashita, S.; Yamamoto, T.; Nakazawa, Y.; Tamura, M.; Kato, R. Gapless spin liquid of an organic triangular compound evidenced by thermodynamic measurements. *Nat. Commun*. **2011**, *2*, 1–6.
9. Watanabe, D.; Yamashita, M.; Tonegawa, S.; Oshima, Y.; Yamamoto, H.M.; Kato, R.; Sheikin, I.; Behnia, K.; Terashima, T.; Uji, S.; Shibauchi, T.; Matsuda, Y. Novel Pauli-paramagnetic quantum phase in a Mott insulator. *Nat. Commun*. **2012**, *3*, 1–6.
10. Yamashita, M.; Nakata, N.; Senshu, Y.; Nagata, M.; Yamamoto, H.M.; Kato, R.; Shibauchi, T.; Matsuda, Y. Highly mobile gapless excitations in a two-dimensional candidate quantum spin liquid. *Science* **2010**, *328*, 1246–1248.
11. Bourgeois-Hope, P.; Laliberté, F.; Lefrançois, E.; Grissonnanche, G.; René de Cotret, S.; Gordon, R.; Kitou, S.; Sawa, H.; Cui, H.; Kato, R.; Taillefer, L.; Doiron-Leyraud, N. Thermal conductivity of the quantum spin liquid candidate EtMe$_3$Sb[Pd(dmit)$_2$]$_2$: No evidence of mobile gapless excitations. *Phys. Rev. X* **2019**, *9*, 041051.
12. Ni, J.M.; Pan, B.L.; Song, B.Q.; Huang, Y. Y.; Zeng, J. Y.; Yu, Y. J.; Cheng, E. J.; Wang, L. S.; Dai, D. Z.; Kato, R.; Li, S. Y. Absence of magnetic thermal conductivity in the quantum spin liquid candidate EtMe$_3$Sb[Pd(dmit)$_2$]$_2$. *Phys. Rev. Lett*. **2019**, *123*, 247204.
13. Yamashita, M. Boundary-limited and glassy-like phonon thermal conduction in EtMe$_3$Sb[Pd(dmit)$_2$]$_2$. *J. Phys. Soc. Jpn*. **2019**, *88*, 083702.
14. Fukuyama, H.; Comment on "Boundary-limited and glassy-like phonon thermal conduction in EtMe$_3$Sb[Pd(dmit)$_2$]$_2$. *J. Phys. Soc. Jpn*. **2020**, *89*, 086001.
15. Yamashita, M. Reply to comment on "Boundary-limited and glassy-like phonon thermal conduction in EtMe$_3$Sb[Pd(dmit)$_2$]$_2$". *J. Phys. Soc. Jpn*. **2020**, *89*, 086002.
16. Yamashita, M.; Sato, Y.; Tominaga, T.; Kasahara, Y.; Kasahara, S.; Cui, H.; Kato, R.; Shibauchi, T.; Matsuda, Y. Presence and absence of itinerant gapless excitations in the quantum spin liquid candidate EtMe$_3$Sb[Pd(dmit)$_2$]$_2$. *Phys. Rev. B* **2020**, *101*, 140407.
17. Yamashita, M. Erratum: Boundary-limited and glassy-like phonon thermal conduction in EtMe$_3$Sb[Pd(dmit)$_2$]$_2$. *J. Phys. Soc. Jpn*. **2020**, *89*, 068001.
18. Bretizer, J.G.; Chou, J-H.; Rauchfuss, T.B. A new synthesis of tetrathiooxalate and its conversion to C$_3$S$_5^{2-}$ and C$_4$S$_6^{2-}$. *Inorg. Chem*. **1998**, *37*, 2080–2082.
19. Crystal Structure Analysis Package, Rigaku and Rigaku Americas (2000-2007). 9009 New Trails Dr. The Woodlands TX 77381 USA.
20. Altomare, A.; Cascarano, G.; Giacovazzo, C.; Guagliardi, A.; Burla, M.C.; Polidori, G.; Camalli, M. SIR92: A program for automatic solution of crystal structures by direct methods. *J. Appl. Cryst*. **1994**, *27*, 435.
21. Sheldrick, G.M. Crystal structure refinement with SHELXL. *Acta Crystallogr. Sect. C*. **2015**, *C71*, 3–8.
22. Kagawa, F.; Sato, T.; Miyagawa, K.; Kanoda, K.; Tokura, Y.; Kobayashi, K.; Kumai, R.; Murakami, Y. Charge-cluster glass in an organic conductor. *Nat. Phys*. **2013**, *9*, 419–422.
23. Ueda, K.; Tsumuraya, T.; Kato, R. Temperature dependence of crystal structures and band parameters in quantum spin liquid β'-EtMe$_3$Sb[Pd(dmit)$_2$]$_2$ and related materials. *Crystals* **2018**, *8*, 138.
24. Fujiyama, S.; Kato, R. Fragmented electronic spins with quantum fluctuations in organic Mott insulators near a quantum spin liquid. *Phys. Rev. Lett*. **2019**, *122*, 147204.
25. Itou, T. (Tokyo University of Science, Tokyo, Japan). Personal communication, 2021.